\documentclass[aps,preprint,epsfig,rotate]{revtex4}
\usepackage{graphicx}
\usepackage{bm}
\usepackage{epsfig}
\input epsf
\begin{document}
\title{On accurate computations of slowly convergent atomic properties in few-electron ions and electron-electron correlations.}

\author{Alexei M. Frolov}
 \email[E--mail address: ]{afrolov@uwo.ca}

\affiliation{Department of Applied Mathematics \\
 University of Western Ontario, London, Ontario N6H 5B7, Canada}

\author{David M. Wardlaw}
 \email[E--mail address: ]{dwardlaw@mun.ca}

\affiliation{Department of Chemistry, Memorial University of Newfoundland, St.John's, 
             Newfoundland and Labrador, A1C 5S7, Canada}

\date{\today}

\begin{abstract}

We discuss an approach to accurate numerical computations of slowly convergent properties in two-electron atoms/ions which include the negatively charged Ps$^{-}$ 
($e^{-} e^{+} e^{-}$) and H$^{-}$ ions, He atom and positively charged, helium-like ions from Li$^{+}$ to Ni$^{26+}$. All these ions are considered in their ground 
$1^{1}S-$state(s). The slowly convergent properties selected in this study include the electron-nulceus $\langle r^{2 k}_{eN} \rangle$ and electron-electron $\langle 
r^{2 k}_{ee} \rangle$ expectation values for $k$ = 2, 3, 4 and 5. 

\noindent 
PACS number(s): 36.10.-k and 36.10.Dr

\noindent 
First version 25.10.2015, Preprint-2015-15/1 (this is 3rd version, 7th of July 2016) [at.phys.], 17 pages.

\end{abstract}

\maketitle

%\noindent \vspace{0.2in}

\section{Introduction}

In this communication we investigate the overall accuracy of representation of the exact wave functions by trial variational functions. This problem is of great interest in the physics of
few-body systems, including few-electron atoms and ions. To formulate the problem let us consider a number of truly correlated two-electron atomic systems, i.e. two-electron atoms and ions. 
Recently, a large number of `highly accurate', or even `precise' methods have been developed to perform bound state computations in such systems. Almost all these methods can produce a very 
accurate (or essentially `exact') numerical value of the total energy $E$ of some selected bound state. In addition to the total energy one also finds that highly accurate methods in 
applications to truly correlated systems allow one to determine a large number of bound state properties, or expectation values. Convergence rates of these expectation values vary between 
very fast and relatively slow, or even very slow. In general, for an arbitrary few-electron atom/ion one always finds a number of expectation values $\langle X \rangle$ which are difficult to 
determine to high and very high accuracy. The sources of such difficulties are mainly related to very slow convergence rates. For instance, for truly correlated atomic systems the expectation 
values of the $p^{2 n}_{e}, r^{2 n}_{eN}$ and $r^{2 n}_{ee}$ operators converge increasingly slowly as the power $2 n$ increases. For large $2 n$ powers, e.g., if $2 n \ge 10$, these expectation 
values converge very slowly. Note that this conclusion is true only for atomic systems for which contribution of the electron-electron correlations is sufficiently large. However, if the role of 
such correlations is small (negligible), then the expectation values of the $p^{2 n}_{e}, r^{2 n}_{eN}$ and $r^{2 n}_{ee}$ operators are simply related with the expectation values of the 
$p^{2}_{e}, r^{2}_{eN}$ and $r^{2}_{ee}$ operators, i.e. the problem can easily be solved by using such `additional' relations. Below, we shall assume that the electron-electron correlations are 
relatively large for all systems mentioned in this study.   
   
The goal of this study is to perform highly accurate computations of some slowly convergent expectation values $\langle r^{2 n}_{eN} \rangle$ and $\langle r^{2 n}_{ee} \rangle$ (where $2 n \ge 6$) 
for a large number of two-electron ions, including the negatively charged Ps$^{-}$ and H$^{-}$ ions, and a number of helium-like atoms and positively charged ions (all ions from Li$^{+}$ to 
Ni$^{26+}$). Due to the electron-electron correlations mentioned above, these expectation values have never been calculated for such ions/atoms to high and very high accuracy. Therefore, our 
expectation values of the $r^{2 n}_{eN}$ and $r^{2 n}_{ee}$ operators can be considered as an important addition to the known bound state properties of all ions/atoms considered in this study. On 
the other hand, the knowledge of these expectation values allows one to understand the internal nature of electron-electron correlations and contribution of these correlations to slowly convergent 
bound state properties. In addition to these two factors, it should be mentioned that the electron-nucleus $\langle r^{2 k}_{eN} \rangle$ expectation value plays an important role in applcations to 
actual physical problems. Indeed, these expectation values are needed to determine electric multipole momenta of different orders in atoms and ions. The same values can be used to determine the 
form-factors of these atoms and ions, i.e. Fourier transformations $\exp(\imath {\bf q}\cdot {\bf r})$ of the one-electron density distribution function $\rho_e({\bf r})$, or one-electron density, 
for short. The rigorous definition of the form-factor is (see, e.g., \cite{Mott}, \cite{Kras}): 
\begin{eqnarray}
  F({\bf q}) = \int \exp(\imath {\bf q}\cdot {\bf r}) \rho_e({\bf r}) d^3{\bf r} = \frac{4 \pi}{q} \int_{0}^{\infty} \sin(q r) \rho_e(r) r dr \label{eq0} 
\end{eqnarray}
where we have assumed that the electron density $\rho_{e}(r)$ is spherically symmetric. For two-electron ions with large electric charges $Q e$ this is a realistic and very accurate approximation. 
In addition, as follows from the general theory of bound state spectra the bound state wave function of atoms is always a real function, or it can be chosen as real. In the case of an ion/atom 
with $N_e$ bound electrons the atomic form-factor (of form-factor, for short) in Eq.(\ref{eq0}) the non-Coulomb part in the last formula is reduced to the form
\begin{eqnarray}
  f(q) = \langle r_{eN}^{2} \rangle - \frac{q^{2} \langle r_{eN}^{4} \rangle}{3!} + \frac{q^{4} \langle r_{eN}^{6} \rangle}{5!} - \frac{q^{6} \langle r_{eN}^{8} \rangle}{7!} + 
  \frac{q^{8} \langle r_{eN}^{10} \rangle}{9!} - \frac{q^{10} \langle r_{eN}^{12} \rangle}{11!} + \ldots \label{ffq}
\end{eqnarray}
where $r = r_{eN}$ is the electron-nucleus distance (scalar coordinate). In Eq.(\ref{ffq}) we eliminated the factor $4 \pi$ which is usually compensated by the corresponding factors from the angular 
part of the total wave function. Here and everywhere below in this study we assume that the electron density $\rho_e(r)$ is normalized to the number of bound electrons $N_e$. 

As follows from Eq.(\ref{ffq}) numerical computations of form-factors for few-electrons atoms and ions are reduced to accurate calculations of a few $\langle r^{2 k}_{eN} \rangle$ expectation values 
($k = 1, 2, 3, \ldots$). In turn, the form-factor determines cross-sections of many actual atomic processes and reactions, e.g., cross-sections of the elastic and non-elastic (electron) scattering, 
bremsstrahlung, positron annihilation, etc (see, e.g., \cite{Kras}, \cite{Beth} and \cite{LLQ}). Numerous examples of calculations of different atomic cross-sections by using form-factors can be
found in Quantum Electrodynamics (see, e.g., \cite{AB} and \cite{Grein}). Formally, this means that, if we know a very few electron-nucleus $\langle r^{2 k}_{eN} \rangle$ expectation values, then we 
can evaluate a significant number of actual atomic properties. In other words, the electron-nucleus $\langle r^{2 k}_{eN} \rangle$ and electron-electron $\langle r^{2 k}_{ee} \rangle$ expectation 
values are important properties which are useful to know for an arbitrary bound state in few- and many-electron atoms and ions. However, by performing accurate atomic calculations it is easy to find 
that the both $\langle r^{2 k}_{eN} \rangle$ and $\langle r^{2 k}_{ee} \rangle$ expectation values become slowly convergent quantities already at $2 k = 6$.     

In this short communication we investigate the problem of slow convergence of these expectation values and determine the $\langle r^{2k}_{eN} \rangle$ and $\langle r^{2k}_{ee} \rangle$ expectation 
values (for $k$ = 3, 4, 5) for a large number of two-electron ions which include the neutral helium atom, the positively charged, helium-like Li$^{+}$ - Ni$^{26+}$ ions ($3 \le Q \le 28$) and the 
weakly-bound, negatively charged Ps$^{-}$ and H$^{-}$ ions, where the convergence of these expectation values is poor. For the last two ions we investigate convergence of these expectation 
values numerically by considering these values as a function of the total number of basis functions $N$. To simplify some parts of our analysis in this study we restrict ourselves to the case of the 
ground (bound) $1^1S-$state in each two-electron ion/atom considered. Another question discussed in this study is the explicit `asymptotic' formulas for the $Q^{-1}$ expansion of these expectation 
values. After an extensive investigation of this probelm we have found that the existing $Q^{-1}$ expansion for the $\langle r^{2k}_{eN} \rangle$ and $\langle r^{2k}_{ee} \rangle$ expectation values 
(for $k$ = 3, 4, 5) needs to be modified. However, after relatively small modifications the $Q^{-1}$-expansion begins to work very well even for slowly convergent properties. Briefly, we can say 
that the accurate determination of the $\langle r^{6}_{eN} \rangle, \langle r^{6}_{ee} \rangle, \langle r^{8}_{eN} \rangle, \langle r^{8}_{ee} \rangle, \langle r^{10}_{eN} \rangle$ and $\langle 
r^{10}_{ee} \rangle$ expectation values in this paper can be considered as an important addition to the atomic properties known for each of the two-electron ions mentioned above.  

\section{Hamiltonian and wave functions}      

The non-relativistic Hamiltonian $H$ of an arbitrary two-electron atom/ion takes the form (see, e.g., \cite{LLQ})
\begin{eqnarray}
 H = -\frac{\hbar^2}{2 m_e} \Bigl[ \nabla^2_1 + \nabla^2_2 + \frac{m_e}{M} \nabla^2_3 \Bigr] - \frac{Q e^2}{r_{32}} - \frac{Q e^2}{r_{31}} + \frac{e^2}{r_{21}} \; \; \; \label{Hamil}
\end{eqnarray}
where $\hbar = \frac{h}{2 \pi}$ is the reduced Planck constant and $m_e$ is the electron mass and $e$ is the absolute value of the electric charge of an electron. In this equation and everywhere 
below in this study the subscripts 1 and 2 designate two electrons ($e^{-}$), while the subscript 3 denotes the positively charged atomic nucleus with the electric charge $Q e$. For all ions/atoms 
considered in this study we assume that the mass of the central atomic nucleus is infinite, i.e. $\frac{m_e}{M} = 0$ in Eq.(\ref{Hamil}). For the Ps$^{-}$ ion the subscript 3 denotes the positron 
($e^{+}$) with the mass $m_{e}$ (the same electron mass) and positive electric charge $+e$, or $e$. The mass ratio in Eq.(\ref{Hamil}) is $\frac{m_e}{M} = 1$ for the Ps$^{-}$ ion. In fact, everywhere 
below, only atomic units $\hbar = 1, \mid e \mid = 1, m_e = 1$ are employed. In these units the explicit form of the Hamiltonian $H$, Eq.(\ref{Hamil}), is simplified to the form
\begin{eqnarray}
 H = -\frac{1}{2} \Bigl[ \nabla^2_1 + \nabla^2_2 + \frac{1}{M} \nabla^2_3 \Bigr] - \frac{Q}{r_{32}} - \frac{Q}{r_{31}} + \frac{1}{r_{21}} \; \; \; \label{Hamil1}
\end{eqnarray}
where the notations $r_{ij} = \mid {\bf r}_i - {\bf r}_j \mid = r_{ji}$ stand for three interparticle distances (= relative coordinates) which are the absolute values of differences of the Cartesian 
coordinates ${\bf r}_i$ of these three particles. Note that each relative coordinate $r_{ij}$ is a scalar which is rotationally and translationally invariant. However, these coordinates are not truly 
independent, since e.g., $\mid r_{32} - r_{31} \mid \le r_{21} \le  r_{32} + r_{31}, \mid r_{32} - r_{21} \mid \le r_{31} \le  r_{32} + r_{21}$, etc. This produces a number of problems in computations 
of the three-particle integrals in the relative coordinates $r_{32}, r_{31}, r_{21}$. To simplify such calculations it is better to apply a set of three perimetric coordinates $u_1, u_2, u_3$ which are 
simply related to the relative coordinates: $u_{k} = \frac12 (r_{ik} + r_{jk} - r_{ij})$, while inverse relations take the form $r_{ij} = u_i + u_j$. The three perimetric coordinates $u_1, u_2, u_3$ are 
truly independent of each other and each of them varies between 0 and $+ \infty$. The Jacobian of the transition $(r_{32}, r_{31}, r_{21}) \rightarrow (u_1, u_2, u_3)$ is a constant which equals 2. 
    
To solve the non-relativistic Schr\"{o}dinger equation $H \Psi = E \Psi$ for the two-electron ions, where $E < 0$, and obtain highly accurate wave function(s) we approximate the unknown exact solution 
of the non-relativistic Schr\"{o}dinger equation by using some rapidly convergent variational expansions. The best of such expansions is the exponential variational expansion in the relative coordinates 
$r_{32}, r_{31}, r_{21}$, or in the perimetric coordinates $u_1, u_2, u_3$. For the ground (bound) $1^1S-$state of the two-electron ions/atoms the explicit form of this expansion is
\begin{eqnarray}
  \Psi &=& \frac12 ( 1 + \hat{P}_{12} ) \sum^{N}_{i=1} C_i \exp(-\alpha_i r_{32} - \beta_i r_{31} - \gamma_i r_{21}) \nonumber \\
  &=& \frac12 ( 1 + \hat{P}_{12} ) \sum^{N}_{i=1} C_i \exp[-(\alpha_{i} + \beta_{i}) u_{3} - (\alpha_{i} + \gamma_{i}) u_{2} - (\beta_{i} + \gamma_{i}) u_{3}] \; \; \; \label{exp}
\end{eqnarray}
where the notation $\hat{P}_{12}$ stands for the permutation operator of identical particles (electrons), $C_i$ ($i = 1, 2, \ldots, N$) are the linear parameters of the exponential expansion, 
Eq.(\ref{exp}), while $\alpha_i, \beta_i$ and $\gamma_i$ are the non-linear parameters of this expansion. The non-linear parameters must be varied in calculations to increase the overall efficiency 
and accuracy of the method. The best-to-date optimization strategy for these non-linear parameters was described in \cite{Fro2001}, while its modified (advanced) version was presented in another paper 
published in 2006 (see the second paper in Ref.\cite{Fro2001}). The $3 N-$conditions $\alpha_{i} + \beta_{i} > 0, \alpha_{i} + \gamma_{i} > 0, \beta_{i} + \gamma_{i} > 0$ for $i = 1, 2, \ldots, N$ must be 
obeyed to guarantee convergence of all three-particle integrals which are needed in computations. 
 
\section{Calculations of the expectation values and three-particle integrlas}

By using highly accurate, variational wave functions $\Psi$ constructed for the ground $1^1S-$state of the two-electron ions we can determine the expectation value of an arbitrary, in principle, 
self-adjoint operator $\hat{X}$ which can be defined for this system. This can be written in the following general form (see, e.g., \cite{Eps})
\begin{equation}
  \langle \hat{X} \rangle = \frac{\langle \Psi \mid \hat{X} \Psi \rangle}{\langle \Psi \mid \Psi \rangle} \label{expt}
\end{equation}
where $\hat{X}$ is a self-adjoint operator which explicitly depends upon three relative coordinates $r_{32}, r_{31}$ and $r_{21}$. Formally, without loss of generality we can assume that our wave 
function has a unit norm, i.e. $\langle \Psi \mid \Psi \rangle = 1$. Our interest in this study is restricted to the cases when in Eq.(\ref{expt}) we choose either $\hat{X} = r^{2k}_{eN}$, or $\hat{X} = 
r^{2k}_{ee}$, where $k$ = 2, 3, 4 and 5. At the beginning of this project we expected that such expectation values could be determined without any problem by using relatively short variational expansions, 
Eq.(\ref{expt}), which include, e.g., $N$ = 700 - 1000 basis functions. However, in actual computations we have found that such relatively short variational expansions of the wave functions allow one to 
determine only two/three correct decimal digits in each of the $\langle r^{8}_{eN} \rangle, \langle r^{10}_{eN} \rangle, \langle r^{8}_{ee} \rangle,$ and $\langle r^{10}_{ee} \rangle$ expectation values. 
After a number of attempts to resolve this situation and improve the overall convergence rate for these expectation values we have decided to investigate this problem more carefully.

First, note that all matrix elements of the $r^{2k}_{eN}$ and $r^{2k}_{ee}$ operators in the exponential basis, Eq.(\ref{exp}), are written in the form of the corresponding Laplace transformations. The 
explicit expressions for these matrix elements in the relative coordinates take the form  
\begin{eqnarray}
 {\cal F}_{2 k + 1;1;1}(\alpha, \beta, \gamma) = \int_0^{+\infty} \int_0^{+\infty} \int_{\mid r_{32} - r_{31} \mid}^{r_{32} + r_{31}} \exp[-\alpha r_{32} - \beta r_{31} - \gamma r_{21}] r^{2k+1}_{32} 
 r_{31} r_{21} dr_{32} dr_{31} dr_{21} \label{e10}
\end{eqnarray}      
where $k$ = 2, 3, 4 and 5. 
These integrals belong to a special class of the general three-body integrals which are the Laplace transforms of the polynomial function $F(r_{32}, r_{31}, r_{21}) = r^{k}_{32} r^{l}_{31} r^{n}_{21}$ 
of the three relative coordinates $r_{32}, r_{31}, r_{21}$. The general formula for such a three-body integral is written in the form
\begin{eqnarray}
  {\cal F}_{k;l;n}(\alpha, \beta, \gamma) = \int \int \int r^{k}_{32} r^{l}_{31} r^{n}_{21} \exp(-\alpha r_{32} - \beta r_{31} - \gamma r_{21}) dr_{32} dr_{31} dr_{21} \label{e101}
\end{eqnarray}
where all indexes $k, l, n$ are assumed to be non-negative. In perimetric coordinates the same integral, Eq.(\ref{e10}), is written in the form
\begin{eqnarray}
  {\cal F}_{k;l;n}(\alpha, \beta, \gamma) = 2 \int_0^{+\infty} \int_0^{+\infty} \int_0^{+\infty} (u_2 + u_3)^{k} (u_1 + u_3)^{l} (u_1 + u_2)^{n} \times \nonumber \\
 \exp[-(\alpha + \beta) u_3 - (\alpha + \gamma) u_2 - (\beta + \gamma) u_1] du_1 du_2 du_3 \label{e11}
\end{eqnarray}
Derivation of the closed analytical formula for the integral, Eq.(\ref{e11}), is straightforward in perimetric coordinates. The formula takes the from
\begin{eqnarray}
 && {\cal F}_{k;l;n}(\alpha, \beta, \gamma) = 2 \sum^{k}_{k_1=0} \sum^{l}_{l_1=0} \sum^{n}_{n_1=0} C^{k}_{k_1} C^{l}_{l_1} C^{n}_{n_1} \frac{(l-l_1+k_1)!}{(\alpha + \beta)^{l-l_1+k_1+1}}
 \frac{(k-k_1+n_1)!}{(\alpha + \gamma)^{k-k_1+n_1+1}} \frac{(n-n_1+l_1)!}{(\beta + \gamma)^{n-n_1+l_1+1}} \nonumber \\
 &=& 2 \cdot k! \cdot l! \cdot n! \sum^{k}_{k_1=0} \sum^{l}_{l_1=0} \sum^{n}_{n_1=0} \frac{C^{k_1}_{n-n_1+k_1} C^{l_1}_{k-k_1+l_1} C^{n_1}_{l-l_1+n_1}}{(\alpha + 
 \beta)^{l-l_1+k_1+1} (\alpha + \gamma)^{k-k_1+n_1+1} (\beta + \gamma)^{n-n_1+l_1+1}} \label{e12}
\end{eqnarray}
where $C^{m}_{M}$ is the number of combinations from $M$ by $m$ (here $m$ and $M$ are integer non-negative numbers). The formula, Eq.(\ref{e12}), can also be written in a few different (but equivalent!) 
forms. The formula, Eq.(\ref{e12}), was derived for the first time in the middle of 1980's \cite{Fro1987}. The formula, Eq.(\ref{e12}), has been used in all calculations of the electron-nucleus 
$\langle r^{2 k}_{eN} \rangle$ and elecron-electron $\langle r^{2 k}_{ee} \rangle$ expectation values performed for this study. 

\section{Discussions and Conclusion}

As mentioned above the $\langle r^{2k}_{eN} \rangle$ and $\langle r^{2k}_{ee} \rangle$ expectation values for $k$ = 2, 3, 4 and 5 have never been determined to high accuracy for almost all two-electron 
ions/atoms discussed in this study. This means that our results from Tables I - V can be considered as an important addition to the bound state properties known for such ions obtained in earlier studies 
(see, e.g, \cite{Fro2015} for the Ps$^{-}$ ion, \cite{Fro2014} for the H$^{-}$ ion and \cite{Fro2015A} for the positively charged, two-electron ions). In general, many of these expectation values for the 
Ps$^{-}$ and H$^{-}$ ions have been predicted numerically since the middle of 1980's (see, e.g., \cite{BD1} and \cite{Ho} for the Ps$^{-}$ ion). An analogous situation existed for the helium atom, and 
for some helium-like, positively charged ions. The goal of this study was to improve the accuracy of numerical predictions of these expectation values. Unfortunately, this simple approach to such a 
comparison of existing calculations to our new calculations does not work, since for the $\langle r^{2 k}_{eN} \rangle$ and $\langle r^{2 k}_{ee} \rangle$ expectation values we cannot compare our 
numerical values with the results from earlier studies (some comparison can be made only with the quantities obtained at the Hartree-Fock and lowest post-HF level). Therefore, in order to demostrate the 
current level of accuracy in this study we have developed and applied a few different tests. The most interesting test is the convergence investigation performed for the $\langle r^{2 k}_{eN} \rangle$ and 
$\langle r^{2 k}_{ee} \rangle$ expectation values. Convergence of the $\langle r^{2 k}_{eN} \rangle$ and $\langle r^{2k}_{ee} \rangle$ expectation values with the total number of basis functions $N$ used 
in actual highly accurate computations in Eq.(\ref{exp}) is shown in Tables I (for $k$ = 2 and 3) and in Table II (for $k$ = 4 and 5) for the Ps$^{-}$ ion. Analogous results for the ${}^{\infty}$H$^{-}$ 
ion look very similar (see Table III). In general, by comparing the expectation values from Tables I, II and III, computed with different numbers of basis functions $N$, we can approximately predict the 
total number of stable decimal digits in the final results.   

Tables IV and V contain numerical values of the $\langle r^{2k}_{eN} \rangle$ and $\langle r^{2k}_{ee} \rangle$ expectation values in $a.u.$ (for $k$ = 3, 4 and 5) determined for a large number of two-electron 
ions in their ground $1^1S-$state(s). As follows from these Tables the actual $Q-$dependence of these expectaion values is smooth and can be investigated in detail. In particular, by using the $\langle r^{2k}_{eN} 
\rangle$ and $\langle r^{2k}_{ee} \rangle$ expectation values from Tables IV - V we can construct accurate asymptotic formulas for the $Q^{-1}$-expansion. The general ideology of the universal method of the
$Q^{-1}$-expansion for few-electron atoms/ions is discussed in \cite{BS} and \cite{Eps}. For our expectation values this `direct' approach does not work well, since the both $\langle r^{2k}_{eN} \rangle$ and 
$\langle r^{2k}_{ee} \rangle$ expectation values vary with $Q$ in a very wide numerical range. Therefore, for these expectation values the classical method of the $Q^{-1}$-expansion must be modified in the 
following way. Instead of the $\langle r^{2k}_{eN} \rangle$ and $\langle r^{2k}_{ee} \rangle$ expectation values we have to use the logarithms of these values, i.e. $\ln \langle r^{2k}_{eN} \rangle$ and $\ln 
\langle r^{2k}_{ee} \rangle$ values which vary between the lower and upper boundaries, when $Q$ changes from unity up to 30 - 40. In reality, these two boundaries are very close to each other.

For these logarithms we can write the regular $Q^{-1}$-expansions, which starts from the first $Q^2$-term 
\begin{eqnarray}
  & & \ln \langle r^{2k}_{eN} \rangle = a_2 Q^2 + a_1 Q + a_0 + b_1 Q^{-1} + b_2 Q^{-2} + b_3 Q^{-3} + b_4 Q^{-4} + \ldots \label{asymp1} \\
  & & \ln \langle r^{2k}_{ee} \rangle = A_2 Q^2 + A_1 Q + A_0 + B_1 Q^{-1} + B_2 Q^{-2} + B_3 Q^{-3} + B_4 Q^{-4} + \ldots \label{asymp2} 
\end{eqnarray}
This structure of the $Q^{-1}$-expansions follows from the explicit form of the Hamiltonian Eq.(\ref{Hamil1}) which describes all two-electron ions/atoms considered in this study. The unknown coefficients in 
the right-hand sides of the equalities Eqs.(\ref{asymp1}) and (\ref{asymp2}) can be obtained by solving a system of linear equations with the known $\ln \langle r^{2k}_{eN} \rangle$ and $\ln \langle r^{2k}_{ee} 
\rangle$ expectation values. In general, the total number of unknown coefficients, e.g., $a_2, a_1, a_0, b_1, b_2, b_3, \ldots$ in Eq.(\ref{asymp1}), is always smaller than the total number of the 
computed expectation values $\ln \langle r^{2k}_{eN} \rangle$ and $\ln \langle r^{2k}_{ee} \rangle$. This means that we have to apply the method of least squares (see, e.g., \cite{Hudson}). After a number of 
trials we have obtained the numerical values of the unknown coefficients in Eqs.(\ref{asymp1}) and (\ref{asymp2}) (for $k$ = 3, 4 and 5). These coefficients can be used in actual evaluations for different 
two-electron atoms/ions. Applications of the formula, Eq.(\ref{asymp1}), to determine the $\langle r^{2k}_{eN} \rangle$ expectation values (for $2 k$ = 6 and 8) are illustrated by the results from Table VI. 
The goal of a series of calculations was to approximate the `unknown' $\langle r^{6}_{eN} \rangle$ and $\langle r^{8}_{eN} \rangle$ expectation values for the K$^{17+}$ ion and compare these quantities with the 
`exact' numerical values obtained in a series of sepate (direct) calculations (see Table IV). Results from Table VI illustrate convergence of intermediate results to the `exact' expectation values when the 
total number of terms $M$ in the $Q^{-1}$-expansion, Eq.(\ref{asymp1}), increases. In Table VI we consider expectation values obtanined for $M$ = 14, 16, 18, 20, 22 and 24. 
  
We have determined the $\langle r^{2k}_{eN} \rangle$ and $\langle r^{2k}_{ee} \rangle$ expectation values ($k$ = 2, 3, 4 and 5) for the ground (bound) $1^1S-$state(s) in a large number of two-electron ions. These
ions include the negatively charged, weakly-bound Ps$^{-}$ and H$^{-}$ ions, He atom and positively charged helium-like ions. Note that most of these expectation values have never been evaluated to high accuracy in 
earlier studies performed for these ions. Investigation of the $\langle r^{2k}_{eN} \rangle$ and $\langle r^{2k}_{ee} \rangle$ expectation values ($k$ = 2, 3, 4 and 5) allows us to understand convergence of these
`slowly convergent' expectation values upon the total number of basis functions $N$ used in calculations. Furthermore, the knowledge of the $\langle r^{2k}_{eN} \rangle$ and $\langle r^{2k}_{ee} \rangle$ expectation 
values can be used to predict the atomic form-factor and many other properties of atomic systems considered in this study. In general, the $\langle r^{6}_{eN} \rangle, \langle r^{6}_{ee} \rangle, \langle r^{8}_{eN} 
\rangle, \langle r^{8}_{ee} \rangle, \langle r^{10}_{eN} \rangle$ and $\langle r^{10}_{ee} \rangle$ expectation values determined to high accuracy in this study can be considered as an important addition to the set 
of known atomic properties for each of two-electron ions considered. Analogous studies must be performed for other bound states in two-electron ions and atoms, e.g., for the $P-$ and $D-$states. For the Ps$^{-}$ and
H$^{-}$ ions the expectation values of the $r^{2k}_{eN}$ and $r^{2k}_{ee}$ operators ($k \ge 3$) are reported for the first time. Accurate $\langle r^{2k}_{eN} \rangle$ and $\langle r^{2k}_{ee} \rangle$ expectation 
(for $k = 3, 4, 5, \ldots$) are important to investigate the role of electron-electron correlations in few-electron atoms and ions. Based on our accurate results we have developed a modified version of the 
$Q^{-1}$-expansion for two-electron ions with different nuclear charges $Q$. This method allows one to predict accurate numerical values of the $\langle r^{6}_{eN} \rangle, \langle r^{6}_{ee} \rangle, \langle 
r^{8}_{eN} \rangle, \langle r^{8}_{ee} \rangle, \langle r^{10}_{eN} \rangle$ and $\langle r^{10}_{ee} \rangle$ expectation values without extensive computations.

\newpage
\begin{table}[tbp]
   \caption{Convergence of the expectation values of the 4th and 6th powers of the electron-nuclear and electron-electron distances for the 
            Ps$^{-}$ ion. All values are shown in atomic units and $N$ is the total number of basis functions used.}
     \begin{center}
%     \scalebox{0.80}{%
     \begin{tabular}{| c | c | c |}
      \hline\hline
 $N$  & $\langle r^{4}_{eN} \rangle$ & $\langle r^{4}_{ee} \rangle$ \\ 
     \hline
 700  & 9.93063866354016104309483$\cdot 10^{4}$ & 2.10544533530693363256539$\cdot 10^{4}$ \\
 1000 & 9.93063867695759270445863$\cdot 10^{4}$ & 2.10544533816002294010497$\cdot 10^{4}$ \\
 1500 & 9.93063867981220599278244$\cdot 10^{4}$ & 2.10544533892937124105159$\cdot 10^{4}$ \\
 2000 & 9.93063867979626166786078$\cdot 10^{4}$ & 2.10544533892588984674548$\cdot 10^{4}$ \\
 2500 & 9.93063867979600602716265$\cdot 10^{4}$ & 2.10544533892583617688472$\cdot 10^{4}$ \\
 3000 & 9.93063867979600416141868$\cdot 10^{4}$ & 2.10544533892583581109197$\cdot 10^{4}$ \\
 3500 & 9.93063867979600415981681$\cdot 10^{4}$ & 2.10544533892583581051440$\cdot 10^{4}$ \\
 3840 & 9.93063867979600415474280$\cdot 10^{4}$ & 2.10544533892583580951959$\cdot 10^{4}$ \\ 
         \hline\hline
 $N$  & $\langle r^{6}_{eN} \rangle$ & $\langle r^{6}_{ee} \rangle$ \\ 
         \hline
 700  & 4.80568068814743249392206$\cdot 10^{6}$ & 9.99929836506863663378819$\cdot 10^{6}$ \\
 1000 & 4.80568114031556600853508$\cdot 10^{6}$ & 9.99929927561165508453627$\cdot 10^{6}$ \\
 1500 & 4.80568125153088967470141$\cdot 10^{6}$ & 9.99929952003726374529628$\cdot 10^{6}$ \\
 2000 & 4.80568125107402346104496$\cdot 10^{6}$ & 9.99929951911244171039228$\cdot 10^{6}$ \\
 2500 & 4.80568125106545530348401$\cdot 10^{6}$ & 9.99929951909487275907597$\cdot 10^{6}$ \\
 3000 & 4.80568125106541274957314$\cdot 10^{6}$ & 9.99929951909478846572168$\cdot 10^{6}$ \\ 
 3500 & 4.80568125106541138515854$\cdot 10^{6}$ & 9.99929951909478594810657$\cdot 10^{6}$ \\ 
 3840 & 4.80568125106541091239665$\cdot 10^{6}$ & 9.99929951909478493674233$\cdot 10^{6}$ \\ 
         \hline \hline
  \end{tabular}
  \end{center}
  \end{table}
\begin{table}[tbp]
   \caption{Convergence of the expectation values of 8th and 10th powers of the electron-nuclear and electron-electron distances for the 
            Ps$^{-}$ ion. All values are shown in atomic units and $N$ is the total number of basis functions used.}
     \begin{center}
%     \scalebox{0.80}{%
     \begin{tabular}{| c | c | c |}
      \hline\hline
 $N$  & $\langle r^{8}_{eN} \rangle$ & $\langle r^{8}_{ee} \rangle$ \\ 
         \hline
 700  & 4.24774492689891131699935$\cdot 10^{9}$ & 8.66821845956715753201073$\cdot 10^{9}$ \\
 1000 & 4.24772820309251553342332$\cdot 10^{9}$ & 8.66819102432451089731615$\cdot 10^{9}$ \\ 
 2000 & 4.24772576076621174183137$\cdot 10^{9}$ & 8.66818742036339759667927$\cdot 10^{9}$ \\
 2500 & 4.24772575007510814637651$\cdot 10^{9}$ & 8.66818739918510066202142$\cdot 10^{9}$ \\
 3000 & 4.24772574985746459705903$\cdot 10^{9}$ & 8.66818739874325727683485$\cdot 10^{9}$ \\
 3500 & 4.24772574985738081866795$\cdot 10^{9}$ & 8.66818739874058038925848$\cdot 10^{9}$ \\
 3840 & 4.24772574985736450997606$\cdot 10^{9}$ & 8.66818739874054267042955$\cdot 10^{9}$ \\
         \hline\hline
 $N$  & $\langle r^{10}_{eN} \rangle$ & $\langle r^{10}_{ee} \rangle$ \\ 
         \hline
 700  & 6.00429250070626982715951$\cdot 10^{12}$ & 1.2133290535975306506086$\cdot 10^{13}$ \\
 1000 & 5.99839628558760769260162$\cdot 10^{12}$ & 1.2124018782302376120116$\cdot 10^{13}$ \\
 1500 & 5.99044448718695409309036$\cdot 10^{12}$ & 1.2108720042936564326664$\cdot 10^{13}$ \\
 2000 & 5.99044264859592495429707$\cdot 10^{12}$ & 1.2108716381259421804620$\cdot 10^{13}$ \\
 2500 & 5.99044264135992752720215$\cdot 10^{12}$ & 1.2108716367982696982857$\cdot 10^{13}$ \\
 3000 & 5.99044264126434938468518$\cdot 10^{12}$ & 1.2108716367785339970708$\cdot 10^{13}$ \\
 3500 & 5.99044264126106870127531$\cdot 10^{12}$ & 1.2108716367779335976792$\cdot 10^{13}$ \\
 3840 & 5.99044264126106995225945$\cdot 10^{12}$ & 1.2108716367779231713846$\cdot 10^{13}$ \\
         \hline \hline
  \end{tabular}
  \end{center} 
  \end{table}
\begin{table}[tbp]
   \caption{Convergence of the expectation values of 4th, 6th, 8th and 10th powers of the electron-nuclear and electron-electron distances 
            for the H$^{-}$ ion with the infinitely heavy nucleus (${}^{\infty}$H$^{-}$). All values are shown in atomic units and $N$ is 
            the total number of basis functions used.}
     \begin{center}
%     \scalebox{0.80}{%
     \begin{tabular}{| c | c | c |}
      \hline\hline
 $N$  & $\langle r^{4}_{eN} \rangle$ & $\langle r^{4}_{ee} \rangle$ \\ 
         \hline
 2500 & 6.45144542412223384151458$\cdot 10^{2}$ & 1.5900946039394936532910$\cdot 10^{3}$ \\
 3000 & 6.45144542412219611111303$\cdot 10^{2}$ & 1.5900946039394858104929$\cdot 10^{3}$ \\
 3500 & 6.45144542412219389586010$\cdot 10^{2}$ & 1.5900946039394853349180$\cdot 10^{3}$ \\
 4000 & 6.45144542412219370980781$\cdot 10^{2}$ & 1.5900946039394852936401$\cdot 10^{3}$ \\
         \hline\hline
 $N$  & $\langle r^{6}_{eN} \rangle$ & $\langle r^{6}_{ee} \rangle$ \\ 
         \hline
 2500 & 8.72661424069925625718951$\cdot 10^{4}$ & 2.1253344237087617700834$\cdot 10^{5}$ \\
 3000 & 8.72661424069612058896071$\cdot 10^{4}$ & 2.1253344237081430766096$\cdot 10^{5}$ \\
 3500 & 8.72661424069594537727044$\cdot 10^{4}$ & 2.1253344237081083393021$\cdot 10^{5}$ \\
 4000 & 8.72661424069592696935431$\cdot 10^{4}$ & 2.1253344237081047127748$\cdot 10^{5}$ \\
         \hline \hline
 $N$  & $\langle r^{8}_{eN} \rangle$ & $\langle r^{8}_{ee} \rangle$ \\ 
         \hline
 2500 & 2.2035718695630030404609$\cdot 10^{7}$ & 5.2218676447807289690830$\cdot 10^{7}$ \\
 3000 & 2.2035718695497230895897$\cdot 10^{7}$ & 5.2218676447555114704017$\cdot 10^{7}$ \\
 3500 & 2.2035718695491884707056$\cdot 10^{7}$ & 5.2218676447545607058237$\cdot 10^{7}$ \\                                                   
 4000 & 2.2035718695491376572405$\cdot 10^{7}$ & 5.2218676447544884570831$\cdot 10^{7}$ \\
         \hline\hline
 $N$  & $\langle r^{10}_{eN} \rangle$ & $\langle r^{10}_{ee} \rangle$ \\ 
         \hline
 2500 & 8.9379707582741862889150$\cdot 10^{9}$ & 2.0815154470686552817302$\cdot 10^{10}$ \\
 3000 & 8.9379707583541128919721$\cdot 10^{9}$ & 2.0815154470904085768025$\cdot 10^{10}$ \\
 3500 & 8.9379707584098506889807$\cdot 10^{9}$ & 2.0815154471021717913050$\cdot 10^{10}$ \\
 4000 & 8.9379707584209025581073$\cdot 10^{9}$ & 2.0815154471047399049048$\cdot 10^{10}$ \\
         \hline \hline
  \end{tabular}
  \end{center}
  \end{table}
\begin{table}[tbp]
   \caption{Expectation values of the $r^{k}_{eN}$ operators for $k$ = 6, 8 and 10 (in $a.u.$) for the ground $1^1S-$state(s) in some two-electron atoms and ions.}
     \begin{center}
     \scalebox{0.95}{%
     \begin{tabular}{| c | c | c | c | c | c |}
      \hline\hline
  ion  & $\langle r^{6}_{eN} \rangle$ & $\langle r^{8}_{eN} \rangle$ & $\langle r^{10}_{eN} \rangle$ \\
          \hline    
   H$^{-}$    &  8.7266142406959270$\cdot 10^{6}$  & 2.20357186954914$\cdot 10^{7}$  & 8.9379707584209$\cdot 10^{9}$ \\ 
   He         &  26.282446975525706                & 289.827674593716                & 4797.7265440809 \\
   Li$^{+}$   &  1.2153694626387516                & 4.57848899403444                & 25.643305753941 \\
   Be$^{2+}$  &  0.16289999058428645               & 0.307966596103007               & 0.86217163108619 \\
   B$^{3+}$   &  3.6352298112380881$\cdot 10^{-2}$ & 4.12490989522882$\cdot 10^{-2}$ & 6.9160629356510$\cdot 10^{-2}$  \\
   C$^{4+}$   &  1.0971961307603750$\cdot 10^{-2}$ & 8.29648580341205$\cdot 10^{-3}$ & 9.2569896501927$\cdot 10^{-3}$ \\
   N$^{5+}$   &  4.0458545441354921$\cdot 10^{-3}$ & 2.18393437236692$\cdot 10^{-3}$ & 1.7378981971464$\cdot 10^{-3}$ \\ 
   O$^{6+}$   &  1.7207164884993150$\cdot 10^{-3}$ & 6.96222590624758$\cdot 10^{-4}$ & 4.1499485730461$\cdot 10^{-4}$ \\ 
   F$^{7+}$   &  8.1438807613848554$\cdot 10^{-4}$ & 2.56150307414319$\cdot 10^{-4}$ & 1.1862759922226$\cdot 10^{-4}$ \\ 
   Ne$^{8+}$  &  4.1882979208164487$\cdot 10^{-4}$ & 1.05338028889520$\cdot 10^{-4}$ & 3.8992426943743$\cdot 10^{-5}$ \\                  
        \hline 
  Na$^{9+}$    & 2.3019815537090110$\cdot 10^{-4}$ & 4.73497099763261$\cdot 10^{-5}$ & 1.4329626527333$\cdot 10^{-5}$ \\
  Mg$^{10+}$   & 1.3358999138069766$\cdot 10^{-4}$ & 2.28900874982039$\cdot 10^{-5}$ & 5.7690341921676$\cdot 10^{-6}$ \\
  Al$^{11+}$   & 8.1118206099732403$\cdot 10^{-5}$ & 1.17570648695508$\cdot 10^{-5}$ & 2.5058841914728$\cdot 10^{-6}$ \\
  Si$^{12+}$   & 5.1181008635663217$\cdot 10^{-5}$ & 6.35647216973809$\cdot 10^{-6}$ & 1.1606982837876$\cdot 10^{-6}$ \\
  Ph$^{13+}$   & 3.3370916846780790$\cdot 10^{-5}$ & 3.59099339836029$\cdot 10^{-6}$ & 5.6804564641147$\cdot 10^{-6}$ \\
  S$^{14+}$    & 2.2386933069629092$\cdot 10^{-5}$ & 2.10740403639127$\cdot 10^{-6}$ & 2.9158091284563$\cdot 10^{-7}$ \\
  Cl$^{15+}$   & 1.5397251950625338$\cdot 10^{-5}$ & 1.27863754695153$\cdot 10^{-6}$ & 1.5604637828244$\cdot 10^{-7}$ \\
  Ar$^{16+}$   & 1.0825341985580918$\cdot 10^{-5}$ & 7.98933602270988$\cdot 10^{-7}$ & 8.6642831006576$\cdot 10^{-8}$ \\
          \hline 
  K$^{17+}$    & 7.7612111364041836$\cdot 10^{-6}$ & 5.12411362175347$\cdot 10^{-7}$ & 4.9707090637754$\cdot 10^{-8}$ \\
  Ca$^{18+}$   & 5.6625739359113182$\cdot 10^{-6}$ & 3.36416344886887$\cdot 10^{-7}$ & 2.9364007261172$\cdot 10^{-8}$ \\
  Sc$^{19+}$   & 4.1969687507078452$\cdot 10^{-6}$ & 2.25564343790269$\cdot 10^{-7}$ & 1.7809620119723$\cdot 10^{-8}$ \\          
  Ti$^{20+}$   & 3.1553271100240321$\cdot 10^{-6}$ & 1.54144738758072$\cdot 10^{-7}$ & 1.1062488796844$\cdot 10^{-8}$ \\
  V$^{21+}$    & 2.4031333585703164$\cdot 10^{-6}$ & 1.07176714358115$\cdot 10^{-7}$ & 7.0223414061394$\cdot 10^{-9}$ \\
  Cr$^{22+}$   & 1.8520276933457352$\cdot 10^{-6}$ & 7.57063449688798$\cdot 10^{-8}$ & 4.5472539112903$\cdot 10^{-8}$ \\
  Mn$^{23+}$   & 1.4428645036187975$\cdot 10^{-6}$ & 5.42565975946396$\cdot 10^{-8}$ & 2.9990536557991$\cdot 10^{-9}$ \\
        \hline
  Fe$^{24+}$   & 1.1353655907960862$\cdot 10^{-6}$ & 3.94056247079358$\cdot 10^{-8}$ & 2.0120181986675$\cdot 10^{-9}$ \\
  Co$^{25+}$   & 9.0166466631135737$\cdot 10^{-7}$ & 2.89737207550905$\cdot 10^{-8}$ & 1.3717427143265$\cdot 10^{-9}$ \\
  Ni$^{26+}$   & 7.2220007860909793$\cdot 10^{-7}$ & 2.15474156619540$\cdot 10^{-8}$ & 9.5023197194145$\cdot 10^{-10}$ \\
    \hline\hline
  \end{tabular}}
  \end{center}
  \end{table}
\begin{table}[tbp]
   \caption{Expectation values of the $r^{k}_{ee}$ operators for $k$ = 6, 8 and 10 (in $a.u.$) for the ground $1^1S-$state(s) in some two-electron atoms and ions.}
     \begin{center}
     \scalebox{0.95}{%
     \begin{tabular}{| c | c | c | c | c | c |}
      \hline\hline
  ion  & $\langle r^{6}_{ee} \rangle$ & $\langle r^{8}_{ee} \rangle$ & $\langle r^{10}_{ee} \rangle$ \\
          \hline    
   H$^{-}$    & 2.1253344237081047$\cdot 10^{5}$  & 5.221867644754489$\cdot 10^{7}$  & 8.9379707584209$\cdot 10^{10}$ \\ 
   He         & 1.1245631448121300$\cdot 10^{2}$  & 1.470413341085919$\cdot 10^{3}$  & 2.7119255221934$\cdot 10^{4}$  \\ 
   Li$^{+}$   & 5.6292546484902171                & 26.91030017485420                & 180.01353891601 \\
   Be$^{2+}$  & 7.7576731467508007$\cdot 10^{-1}$ & 1.915698537380722                & 6.6096105596883 \\ 
   B$^{3+}$   & 1.7548284756742420$\cdot 10^{-1}$ & 2.642503536372184$\cdot 10^{-1}$ & 5.5571366091173$\cdot 10^{-1}$ \\ 
   C$^{4+}$   & 5.3383732025165301$\cdot 10^{-2}$ & 5.410677232398524$\cdot 10^{-2}$ & 7.6573683711407$\cdot 10^{-2}$ \\
   N$^{5+}$   & 1.9785477877402386$\cdot 10^{-2}$ & 1.441400250285469$\cdot 10^{-2}$ & 1.4661687535537$\cdot 10^{-2}$ \\
   O$^{6+}$   & 8.4446682997834888$\cdot 10^{-3}$ & 4.634451576888273$\cdot 10^{-3}$ & 3.5511567156249$\cdot 10^{-3}$ \\   
   F$^{7+}$   & 4.0071008219829092$\cdot 10^{-3}$ & 1.716011552310627$\cdot 10^{-3}$ & 1.0260492148249$\cdot 10^{-3}$ \\
   Ne$^{8+}$  & 2.0648758933294060$\cdot 10^{-3}$ & 7.091935926647552$\cdot 10^{-4}$ & 3.4009437094400$\cdot 10^{-4}$ \\                  
        \hline 
  Na$^{9+}$   & 1.1366636836485451$\cdot 10^{-3}$ & 3.200492444138807$\cdot 10^{-4}$ & 1.2582662493278$\cdot 10^{-4}$ \\ 
  Mg$^{10+}$  & 6.6046023981302165$\cdot 10^{-4}$ & 1.552212566637046$\cdot 10^{-4}$ & 5.0937094007657$\cdot 10^{-5}$ \\
  Al$^{11+}$  & 4.0145512771582814$\cdot 10^{-4}$ & 7.994117953375549$\cdot 10^{-5}$ & 2.2227468708512$\cdot 10^{-5}$ \\
  Si$^{12+}$  & 2.5351297160481673$\cdot 10^{-4}$ & 4.331857470689201$\cdot 10^{-5}$ & 1.0335731248035$\cdot 10^{-5}$ \\
  Ph$^{13+}$  & 1.6541519866000751$\cdot 10^{-4}$ & 2.451978179637930$\cdot 10^{-5}$ & 5.0752523190550$\cdot 10^{-6}$ \\
  S$^{14+}$   & 1.1103834367892564$\cdot 10^{-4}$ & 1.441385237818165$\cdot 10^{-5}$ & 2.6127229815422$\cdot 10^{-6}$ \\
  Cl$^{15+}$  & 7.6411142361518097$\cdot 10^{-5}$ & 8.758258104512842$\cdot 10^{-6}$ & 1.4018157968702$\cdot 10^{-6}$ \\
  Ar$^{16+}$  & 5.3747821872842266$\cdot 10^{-5}$ & 5.479528022775409$\cdot 10^{-6}$ & 7.8008737661896$\cdot 10^{-7}$ \\
          \hline 
  K$^{17+}$   & 3.8550525434837112$\cdot 10^{-5}$ & 3.518441099086186$\cdot 10^{-6}$ & 4.4842890525060$\cdot 10^{-7}$ \\
  Ca$^{18+}$  & 2.8136886884911498$\cdot 10^{-5}$ & 2.312355602765375$\cdot 10^{-6}$ & 2.6537638458779$\cdot 10^{-7}$ \\ 
  Sc$^{19+}$  & 2.0861340806322627$\cdot 10^{-5}$ & 1.551847129149528$\cdot 10^{-6}$ & 1.6120980250305$\cdot 10^{-7}$ \\          
  Ti$^{20+}$  & 1.5688474182661537$\cdot 10^{-5}$ & 1.061376785918574$\cdot 10^{-6}$ & 1.0027749296477$\cdot 10^{-7}$ \\ 
  V$^{21+}$   & 1.1951752839024337$\cdot 10^{-5}$ & 7.385339159464128$\cdot 10^{-7}$ & 6.3733691984626$\cdot 10^{-8}$ \\ 
  Cr$^{22+}$  & 9.2131446340301449$\cdot 10^{-6}$ & 5.220384819335957$\cdot 10^{-7}$ & 4.1312674172434$\cdot 10^{-8}$ \\ 
  Mn$^{23+}$  & 7.1793199944813362$\cdot 10^{-6}$ & 3.743671833575214$\cdot 10^{-7}$ & 2.7267628879708$\cdot 10^{-8}$ \\ 
        \hline
  Fe$^{24+}$  & 5.6504452814749105$\cdot 10^{-6}$ & 2.720546821859721$\cdot 10^{-7}$ & 1.8300150240156$\cdot 10^{-8}$ \\ 
  Co$^{25+}$  & 4.4882185523678996$\cdot 10^{-6}$ & 2.001406932108432$\cdot 10^{-7}$ & 1.2473265480770$\cdot 10^{-8}$ \\ 
  Ni$^{26+}$  & 3.5955234289593961$\cdot 10^{-6}$ & 1.489162219104389$\cdot 10^{-7}$ & 8.6255426406327$\cdot 10^{-9}$ \\ 
    \hline\hline
  \end{tabular}}
  \end{center}
  \end{table}
\begin{table}[tbp]
   \caption{Convergence of the $\langle r^{6}_{eN} \rangle$ and $\langle r^{8}_{eN} \rangle$ expectation values for the 
            two-electron K$^{17+}$ ion determined with the use of the $M$-terms in the $Q^{-1}$-expansion, Eq.(\ref{asymp1}).}
     \begin{center}
%     \scalebox{0.80}{%
     \begin{tabular}{| c | c | c |}
      \hline\hline
 $M$  & $\langle r^{6}_{eN} \rangle$ & $\langle r^{8}_{eN} \rangle$ \\ 
         \hline
      14 &  7.76122879758595388E-06 & 5.12412920398821052E-07 \\
      16 &  7.76121135024542466E-06 & 5.12411380495447621E-07 \\
      18 &  7.76121108013487460E-06 & 5.12411356758053888E-07 \\
      20 &  7.76121113844470425E-06 & 5.12411362696037344E-07 \\
      22 &  7.76121113636655254E-06 & 5.12411362053147883E-07 \\
      24 &  7.76121113640500495E-06 & 5.12411362204730508E-07 \\
       \hline 
 `exact'$^{(a)}$ &  7.7612111364041836$\cdot 10^{-6}$ & 5.12411362175347$\cdot 10^{-7}$ \\
         \hline\hline
  \end{tabular}
  \end{center}
  ${}^{(a)}$See Table IV.
  \end{table}
\end{document}